# A Patient-Specific Respiratory Model of Anatomical Motion for Radiation treatment Planning

Qinghui Zhang, Ph.D.[1], Alex Pevsner, Ph.D[1], Agung Hertanto, Ph.D[1], Yu-Chi Hu, MS.[1], Kenneth E. Rosenzweig, M.D.[2], C. Clifton Ling, Ph.D.[1], Gig S Mageras, Ph.D.[1]

Departments of [1]Medical Physics and [2]Radiation Oncology, Memorial Sloan-Kettering Cancer Center, New York, NY

**Abstract—Objective**: Modeling of respiratory motion is important for a more accurate understanding and accounting of its effect on dose to cancers in the thorax and abdomen by radiotherapy. We have developed a model of respiration-induced organ motion in the thorax, without the commonly adopted assumption of repeatable breath cycles.

***Methods and Results***: The model describes the motion of a volume of interest within the patient, based on a reference 3-dimensional image (at end-expiration), and the diaphragm positions at different time points. The input data are respiration-correlated CT (RCCT) images of patients treated for nonsmall cell lung cancer, consisting of 3D images, including the diaphragm positions, at 10 phases of the respiratory cycle. A deformable image registration algorithm calculates the deformation field that maps each 3D image to the reference 3D image. A principle component analysis (PCA) is performed to parameterize the 3D deformation field in terms of the diaphragm motion. We show that the first two principal components are adequate to accurately and completely describe the organ motion in the data of 4 patients. Artifacts in the RCCT images that commonly occur at the mid-respiration states are reduced in the model-generated images. Further validation of the model is demonstrated in the successful application of the parameterized 3D deformation field to RCCT data of the same patient but acquired several days later.

***Conclusions***: We have developed a method for predicting respiration-induced organ motion in patients that has potential for improving the accuracy of dose calculation in radiotherapy. Possible limitations of the model are cases where the correlation between lung tumor and diaphragm position is less reliable such as superiorly situated tumors, and interfraction changes in tumor-diaphragm correlation. The limited number of clinical cases examined suggests but does not confirm the model's applicability to a wide range of patients.

Manuscript received March 9, 2006. This work is supported by XXX.

## I. INTRODUCTION

R espiration-induced anatomic motion can limit the accuracy of dose calculation and delivery in radiotherapy of cancers in the thorax and abdomen. To improve high-precision conformal radiotherapy and intensity-modulated radiotherapy (IMRT) of these disease sites one needs to more precisely understand respiration-induced anatomical motion and account for its effect. To effect such improvements, two types of studies have been conducted: 1) to measure the positions of the tumor and organs-at-risk (OARs) at multiple phases in the respiratory cycle using respiration-correlated methods [1-8], and 2) to estimate the effect of respiration-induced motion on the delivered dose to the tumor and OARs [9-19].

In evaluating item 2) above, most groups assume that patients breathe regularly, in spite of evidence to the contrary [20-22]. Some algorithms characterize organ motion by a periodic function as measured by certain surrogates of the breathing cycle (e.g. the Varian RPM system); however, many patients have irregular breathing such that a periodic function is inadequate to describe the organ motion [20-22]. Variability in respiration during treatment can be measured with fluoroscopy or respiration monitors, but they do not provide adequate information on the internal 3D motion. As to the possible use of a respiratory correlated computed tomography (RCCT) image set to calibrate respiration monitors, interfractional variations in internal anatomical positions would introduce uncertainties [23-27].

Recognizing the potential benefit of models of organ motion that do not assume reproducible respiration patterns, Low *et al* described a breathing motion model parameterized by tidal volume and airflow measured with spirometry, allowing characterization of hysteresis and irregular breathing patterns [28]. Their method uses manual segmentation of image features, thus can provide motion trajectories for only a limited number of anatomical points. Zeng *et al* assumed synthetic periodic respiratory motion functions and derived a 2D motion model in the thorax using projection images from cone-beam CT scans [29]. More recently, the method has been extended to 3D motion [30]. Sohn *et al* [31] used a principal component analysis to model the inter-fractional organ deformation in the male pelvis, and reported that four principal components can accurately describe deformation of bladder, rectum, prostate and seminal vesicles. However, they did not study the relation between deformation and patient-specific surrogates.

The objective of the present work is to estimate thoracic tumor and normal lung motion during treatment without the assumption of repeatable breathing pattern. Our model describes the 3D trajectories of all voxels in a volume of interest, which are parameterized by the motion of a surrogate, taken to be the CT coordinate along the patient longitudinal axis of the most superior point on the diaphragm in this study. As validation, we evaluate the accuracy of the model by applying it to predict anatomic changes in RCCT images of lung cancer patients obtained on different days. Our results indicate that the proposed method is accurate in describing organ motion and potentially useful for improving the accuracy of dose calculation and delivery for cancers in the thorax and abdomen by radiotherapy.

## II. METHODS

Before giving a detailed mathematical account of the methods of deformable object registration, principal component analysis, and the application of these to model respiration-induced organ motion, we shall briefly outline the overall approach of this study.

We start with a motion model that has been developed for free-breathing coronary MR angiography [32] and modify it to incorporate high-dimensional deformations. The earlier study characterized heart motion as an affine transformation that varied according to motion of the diaphragm. A principal components analysis was used to establish the relation between affine motion of the heart and diaphragm motion. The resultant model provided a prediction of the position and shape of the heart according to the position of the diaphragm at one or more time points. In our model, we consider high-dimensional deformations at the CT voxel level, in order to accommodate complex respiratory movement in the thorax. The basic concept of our approach is: 1) adaptation of the model to a particular patient using RCCT images; and 2) application of the improved model to predict 3D deformation in the patient's anatomy.

Four patients treated for lung cancer and consenting to an institutional review board protocol receive a respiration-correlated CT (RCCT) scan [5]. Briefly, a 4-slice scanner (LightSpeed GX/i, GE Medical Systems) acquires repeat CT images for a complete respiratory cycle at each couch position while recording patient respiration (Real-time Position Management [RPM] System, Varian Medical Systems). The CT images are retrospectively sorted (GE Advantage 4D) to produce a series of 3D images $I(\vec{x}, t)$ at 10 respiratory time points, where $I(\vec{x}, t)$ denotes the intensity of voxel at 3D position $\vec{x}$, and $t$ the time point. The time resolution of each CT slice (i.e., gantry rotation period) is 0.5 s; CT slice thickness is 2.5 mm. On a different day, also before the patient has received any radiation treatment, patients receive a helical RCCT on a single-slice scanner (PQ5000, Philips/Marconi Medical Systems). The images are retrospectively sorted into 4 respiratory time points, with 1 s time resolution and 5 mm slice thickness [1].

To generate gross tumor volume (GTV) contours in the lung, the radiation oncologist (K.R.) first outlined the GTV (lung-tumor boundaries) in the end expiration images from the GE scanner. Using that as a visual guide, three other observers independently delineated the GTV on the end inspiration images. This provided a reference to which the model- predicted contours were compared. The intent of having all observers use the same physician-delineated GTV as a guide was to reduce interobserver variations to a level comparable to intraobserver variations (variations for the same observer). In addition, one observer delineated the GTV in all the images of the RCCT scans (from both studies using the GE and the Philips scanners).

To perform deformable registration between the RCCT images at different respiratory phases, the end expiration (EE) image is chosen as the reference, denoted by $I(\vec{x}, t_{ref})$. Deformable registration is applied between each image $I(\vec{x},t)$ at time $t$ and the reference image to obtain a set of deformation fields $\vec{h}(\vec{x},t)$, each of which defines a voxel-dependent displacement field $\vec{u}(\vec{x},t)$, such that $\vec{h}(\vec{x},t) = \vec{x} + \vec{u}(\vec{x},t)$. The point $\vec{h}(\vec{x},t)$ in the reference image corresponds to point $\vec{x}$ in the study image. That is,

$$I(\vec{h}(\vec{x},t), t_{ref}) \rightarrow I(\vec{x},t) \qquad (1).$$

The set of displacement fields $\vec{u}(\vec{x},t)$ deforms the reference image at EE into the images at the other 9 time points, and represents the motion behavior of each tissue voxel over a respiratory cycle.

A major task of the analysis is to connect the surrogate signals with the model parameters given by the time-varying displacement fields:

$$u = Bs$$

(2)

where $u(t) = [\vec{u}_1(t), ..., \vec{u}_m(t), ..., \vec{u}_M(t)]^T$ is the displacement field at time $t$ for all voxels in the respiratory cycle. Here M is the total number of voxels.

$s(t) = [s_1(t), ..., s_N(t)]^T$ is the corresponding vector of two or more surrogate positions relative to the couch. Here N is the number of surrogate signals. In this paper, N=2 and the two surrogate signals are the top of the diaphragm along the patient longitudinal axis in the CT coordinate system and its precursor (described below). Knowing $\vec{u}_m(t)$ from the registration and $s(t)$ from measurements, we can determine matrix $B$ (its size is $M \times N$). We note that the many elements of the matrix $B$ will have identical values, i.e. the relation between $u(t)$ and $s(t)$ are the same for many voxels. In addition, there may be inconsistencies in the trajectories of individual voxels from one phase to another, resulting in irregularities in the time-dependent variation of $u$, caused by ambiguities in the deformable registration process. To circumvent this problem, we perform a principal component analysis (PCA) to reduce the complex data set to a lower dimension, by removing noise and redundancy in the model parameters, thereby reducing the inconsistencies between the different deformation fields. It does so by finding a linear combination of the original variables to produce a set of new variables, called principal components. Only a few of the principal components are sufficient to express the important dynamics of the data. We assume that those principal components with the largest variances determine the main motion of the lung, while principal components with small variances are noise coming from the imperfect deformable registration.

*a) Deformable registration algorithm*

For our analysis, we use a fast free-form deformable registration algorithm [33], which minimizes the following energy function

$$E(u) = \int (I_B(\vec{x}+\vec{u}) - I_A(\vec{x}))^2 \, d\vec{x} + \lambda \sum_{i=1}^{3} \int |\nabla u_i|^2 \, d\vec{x}$$

(3)

Here $I_A(\vec{x})$ and $I_B(\vec{x})$ are the intensity of images A and B at point $\vec{x}$.    The first term of Eq. (3) describes the similarity between images A and B, while the second term is a smoothing term that limits sharp gradients in the vector displacement field $\vec{u}$. The adjustable parameter $\lambda$ that controls the relative importance of the two terms is set to a value of 0.1. Taking the variation of Eq. (3) yields the Euler-Lagrange equation

$$\lambda \Delta \vec{u} + (I_B(\vec{x}+\vec{u}) - I_A(\vec{x})) \frac{\partial I_B(\vec{x}+\vec{u})}{\partial \vec{u}} = 0$$

with the following boundary condition:

$$\delta \tilde{u}_S \bullet \frac{\partial \vec{u}}{\partial \hat{n}} = 0.$$

(5)

Here S is the surface of the images, $\hat{n}$ is the normalized normal of the boundary of image, and $\delta \tilde{u}_s$ is the variation of $\vec{u}$ at the boundary of the image. In our implementation, we take $\delta \tilde{u}_s = 0$. That is, we assume that the boundary of image B does not move. Solution of Eq. (4) uses a multiresolution Newton iterative scheme and Gauss-Seidel method [33, 34, 35].

*b) Model adaptation using principal components analysis*

In the principal component analysis, we first construct vectors with displacements of each voxel as its components for each time point. Then we construct centered vectors which are the differences between the constructed vectors and their average. A covariance matrix is constructed and the eigenvectors of the covariance matrix are calculated. Those eigenvectors with the largest eigenvalues are the principal components used in the model.

We describe each of these steps in more detail.  The set of displacement vectors is given by

$$p_j = \left[ u_{1,1,j}, u_{1,2,j}, u_{1,3,j}, ..., u_{M,3,j}, s_{1,j}, s_{2,j}, ..., s_{N,j} \right]^T.$$

(4)

Here $u_{m,i,j}$ is the *i-th* component (i=1-3) of displacement for the voxel *m* at time point *j*, and $s_{n,j}$ is the displacement of the *n-th* surrogate at time point *j*. *M* is the total number of voxels which has a typical value around 9 million, and *N* is the total number of surrogates.  Next we construct a matrix   $P = [\tilde{p}_1, \tilde{p}_2, ..., \tilde{p}_j, ..., \tilde{p}_J]$, composed of centered vectors $\tilde{p}_j = p_j - \overline{p}$ where the mean parameter vector $\overline{p} = \frac{1}{J} \sum_{j=1}^{J} p_j$  represents the respiration averaged motion state.  The size of matrix $P$ is $(3M+N) \times J$.  The covariance matrix $PP^T$ is positive semi-definite, meaning its eigenvalues are non-negative.  Because the size of $PP^T$ is $(3M+N) \times (3M+N)$, determination of its eigenvectors directly is computationally prohibitive.  On the other hand, since $\sum_{j=1}^{J} \widetilde{p}_j = 0$ by definition, we have only J-1 (J=10 for our case) independent measurements to describe those data. Because of this, there are at most J-1 eigenvectors having nonzero eigenvalues [36]. Suppose that $P^T P$ has an eigenvector X with eigenvalue $\lambda$. Multiplying $PP^T$ by $PX$ we obtain $PP^T(PX) = P(P^T PX) = \lambda PX$; thus, $PX$ is an eigenvector of the covariance matrix $PP^T$ and $\lambda$ is a corresponding eigenvalue. Furthermore, $tr(P^T P) = tr(PP^T) = \sum_i \lambda_i$ where $tr$ denotes the trace operation and $\lambda_i$ is an eigenvalue of $P^T P$ or $PP^T$. From this we make the following observations: First, all eigenvalues of the covariance matrix $PP^T$ are zero except those which are nonzero eigenvalues of $P^T P$. Second, we can calculate eigenvectors and eigenvalues of $P^T P$ (which is a $10 \times 10$ matrix in our case), then obtain the corresponding eigenvectors (whose eigenvalues are nonzero) $E = PX$ of the covariance matrix, $PP^T$, of which there are at most 9. We hypothesize that a good approximation of each possible motion state $p(t)$ at an arbitrary time point $t$ can be expressed as a weighted sum of the $K$ eigenvectors $e_k$ with the largest eigenvalues:

$$p(t) \approx \overline{p} + \sum_{k=1}^{K} w_k(t) e_k . \qquad (6)$$

*c) Relating surrogate signals to model parameters*

Instead of expressing a patient's motion state in terms of unknown weighting factors as in Eq. (6), we wish to express it in terms of the surrogate signals. Rewriting Eq. (6) in terms of centered vectors, and omitting the dependence on $t$ for notational simplicity, yields $\widetilde{p} \approx \sum_{k=1}^{K} w_k e_k$, or in matrix notation, $\widetilde{p} \approx EW$ where matrix $E = [e_1 ..., e_K]$ consists of the first $K$ eigenvectors used to approximate the motion states and $W = [..., w_k ,...]^T$ with size $K \times 1$. Noting that $\widetilde{p} = [\widetilde{u} , \widetilde{s}]^T$ where $\widetilde{u} = [\widetilde{u}_{1,1},...,\widetilde{u}_{3,M}]^T$ and $\widetilde{s} = [\widetilde{s}_1 ,...,\widetilde{s}_N]^T$ this system of equations can be split into two separate ones as [32]

$$\widetilde{u} \approx E_u W$$
$$\widetilde{s} \approx E_s W \qquad (7)$$

where $E_u$ and $E_s$ are constructed from the upper *3M* rows and lower $N$ rows of *E*, respectively, for *3M* voxel displacements and $N$ surrogate signals. Eliminating *W* from the equations and assuming an inverse matrix, $E_s^{-1}$ exists yields

$$\widetilde{u}(t) \approx E_u E_s^{-1} \widetilde{s}(t) \equiv B\widetilde{s}(t)$$

(8)

thus establishing the desired relation between displacement field and surrogate signals. The assumption that $E_s^{-1}$ exists implies that the number of rows $N$ in $E_s$ must be equal to or larger than the number of columns $K$; thus there must be least one surrogate signal for each eigenvector used in Eq. (6).

Each 3D image in the series is tagged by two surrogate signals that are visible in the 3D images. We have chosen the combination of current diaphragm position *s(t)* plus its precursor position *s(t-3)* as surrogate measurements, which has the advantage that it incorporates temporal correlations into the model, that is, it distinguishes between the inspiration and expiration portions of the respiratory cycle. The surrogate signals in Eq. (8) are given by

$$s(t) = [d_t, d_{t-3}]^T \qquad t > 4$$
$$\quad = [d_t, d_{t+J-3}]^T \qquad t < 4 \qquad , \qquad (9)$$

where *t* is the respiratory time index from 1 to J and $d_t$ measures the top of the diaphragm along patient longitudinal axis in the CT coordinate system. We estimate the uncertainty in diaphragm position is one half the slice thickness of 2.5mm.

*d) Model application and evaluation*

To apply the methods described above to a patient's organ motion, the surrogate signals at time point *t*, given by Eq. (9), are substituted into Eq. (8), where the two eigenvectors with largest eigenvalues are used in calculating matrix *B*, to obtain the centered displacement field $\widetilde{u}(t)$. Adding to this the respiration-averaged displacement field $\overline{u}$ yields the displacement field *u(t)*, which is applied to deform the reference EE CT image.

We quantify differences between deformed and actual images at a given time point by comparing gross tumor volume (GTV) and lung delineations between the two images. Starting with the GTV contours at the EE time point as a reference, we generate a GTV surface by using the triangulation method given in [37]. According to the displacement field $\vec{u}(\vec{x},t)$, we can predict the new positions of the triangle vertexes at a given time point *t*, thus deforming the GTV. We then reslice the deformed GTV along the axial image planes, yielding a new set of contours corresponding to the deformed image. For the lung contour, we resegment the deformed image at the given time point, using a threshold technique [38].

One test of model accuracy compares GTV centroid position between model-deformed and actual images at various time points in the same RCCT imaging session. In two patients, evaluation of differences between model-predicted and observer-drawn GTV surfaces at end inspiration (EI), as well as interobserver differences in delineated GTV, uses a contour comparison algorithm written for this purpose, which determines the distance between the two surfaces along different directions [39]. This examines whether any GTV shape deformations are accurately modeled. The differences between the two surfaces are sampled along different polar angles (θ = 0°, 30°,.., 180° where 0° and 180° correspond to the superior and inferior directions, respectively), and azimuthal angles (φ = 0°, 36°, 72°,.., 324° where 90° and 270° correspond to posterior and anterior directions) and are plotted as a 2D maps, similar to Mercator maps of the world.

We also investigate the model's ability to predict motion states in a different session, for example, to determine whether the model is applicable at a treatment session subsequent to the simulation session. This is accomplished by comparing model-predicted with actual GTV position in the second RCCT scan on a different day. The two RCCT scans are registered by alignment of the skeletal anatomy, and the diaphragm positions in the second RCCT are used as surrogate signal input to the model calibrated on the first RCCT. In this fashion, model predictions of images and GTV delineations are generated at EE and EI for comparison to the actual images and delineations in the second RCCT.

### III. RESULTS

*a) Spectrum of PCA eigenvalues*

Figure 1 shows the spectrum of eigenvalues for one patient. One can see that the first two eigenvalues accounts for approximately 83% of the cumulative sum of eigenvalues of the system. For the other three patients, the first two eigenvalues accounts respectively for approximately 83%, 90% and 89% of the cumulative sum of system eigenvalues.

*b) Model prediction within a single RCCT session*

We investigate whether two eigenvectors with the largest eigenvalues (i.e., two principal components) are sufficient to accurately describe a patient's motion states within a single RCCT session. Left column in Fig.2 shows red-blue overlays of actual images at end expiration (EE) and end inspiration (EI). Areas of red and blue indicate tissue differences between the two images. Differences between the EE and EI states are clearly visible in the areas of diaphragm, mediastinum and tumor. The second column shows red-blue overlays of the model reconstructed and actual images at EI. Areas of density mismatch are almost entirely eliminated, indicating that the model, based on only 2 principal components, accurately predicts the 3-D changes from EE to EI. Columns 3 and 4 show overlays of PCA reconstructed images and actual CT images at mid-inspiration (MI) and mid-expiration (ME), respectively, again confirming the good agreement between model and actual images and demonstrating that the model accurately predicts changes throughout the respiratory cycle.

We quantify the model's accuracy by applying it to deform the GTV delineations from EE to the other motion states, and comparing it to observer delineations. Figure 3 shows the discrepancy between model-predicted and observed GTV centroid positions in the four patients, at the EI, MI, and ME motion states. Mean discrepancy is less than 1 mm in the left-right (LR) and anterior-posterior (AP) directions, and about 1 mm in the superior-inferior (SI) direction with maximum discrepancy being less than 2 mm. The slightly larger SI discrepancy is likely a result of the coarser CT resolution in that direction (due to the 2.5 mm thick CT slices). To investigate whether the model accurately predicts the GTV shape, we analyze the distance between model-predicted and observer drawn GTV surfaces. Figure 4 shows composite 2D polar maps of the differences between observers, and between the model and an observer differences, for two patients at end inspiration. Figures 4A and 4C show the average model-observer differences in the GTV at EI (4 comparisons), while figures 4B and 4D show average interobserver differences (6 comparisons). The 95% confidence limit (CL) differences (3D vector length) for 4A, 4B, 4C and 4D are 0.22cm, 0.26cm, 0.32cm and 0.21 cm. One can see that model-observer differences are comparable in value to interobserver differences.

Figure 5 compares the 3D vector length displacement of lung voxels from EE to EI, before and after model application. The shape of the lung is the segmentation results of the actual CT image at EE. The left panel shows the predicted displacement of the deformable image registration prior to model application, while the middle panel is that from the motion model using two principal components. The similarity between the two figures is very clear. In addition, one can see that the largest movement occurs near the diaphragm and diminishes with more superior locations. The right panel plots the differences in voxel displacements between the model prediction and deformable registration. It is clear that most differences are around 2-3mm and are a result of some noise removal by the model and the two eigenvalues approximation of the model. Note that the RCCT voxel size is $1 \times 1 \times 2.5$ mm$^3$; therefore, the differences in displacement correspond to about one to two voxels.

We further evaluate the accuracy of using two principal components by comparing model prediction of changes in lung shape with the actual shapes observed in the RCCT images. The left panel of Fig.6 plots the 90% confidence limit (CL) differences (3D vector length) between model-predicted and actual delineated lung surfaces for the different motion states. In most cases the 90% CL differences are 2 mm or less; the largest discrepancy is 3mm, at EI of patient 2. The right panel of Fig. 6 shows the 2D surface difference map for this case. The mean discrepancy is 1 mm, with the largest discrepancy of 4 mm occurring in the superior region of the lung. This is not surprising since, in this study, we have used the diaphragm as our surrogate, which is less likely to accurately model changes in the superior region.

It is well established that RCCT images often exhibit discontinuity artifacts across neighboring CT slices, particularly at rapidly varying mid-respiratory motion states, which are a result of limitations in the retrospective image sorting process [3, 8]. The left column of Figure 7 shows coronal RCCT images of 3 patients at mid-inspiration. Artifacts are clearly visible. Right column of Figure 7 shows the model-generated images at the same respiration state, illustrating that the artifacts are greatly reduced. This result supports the hypothesis, stated earlier, that imperfections in the deformation field can be removed by retaining only the largest principal components. It further suggests an application of the model to remove artifacts in RCCT images.

*c) Model prediction for different sessions*

We examine the ability of the model to predict a patient's motion state on a different day from the one on which the model was calibrated. This is done by means of a second RCCT acquired on a different day, which is registered to the first RCCT by alignment of the vertebral bodies. The diaphragm positions in the second RCCT are used as surrogate signals to the model. Figure 8 shows axial and coronal overlays of the model-predicted deformed image at end inspiration (red) and the actual image from the second RCCT (blue), for two patients. Visually one observes good spatial agreement between the model-predicted and actual GTV that is consistent with the 5 mm slice spacing in the second RCCT. The blue contour is the observer drawn GTV in the second RCCT. Figure 9 compares predicted and actual GTV centroid displacements in the second RCCT, relative to the EE position in the first RCCT, for the four patients. Mean ± one standard deviation discrepancies between model-predicted and observed GTV centroid displacements over the 4 patients and 2 motion states (EE and EI) are 1.1 ± 0.6 mm LR, 1.8 ± 1.0 mm AP, and 1.6 ± 1.4 mm SI. Note in particular that patient 4 shows a large change in GTV position between sessions (1 cm AP, 1.5 cm SI), even at the same motion state (EE, upper row), and that the motion model is able to accurately predict this change. The results suggest that, at least for anatomy in proximity to the diaphragm, a patient-specific motion model derived from an RCCT at simulation may be applicable to a patient's motion states at treatment.

## IV.  DISCUSSION AND CONCLUSIONS

Modeling of respiration-induced motion is important for a more accurate understanding and accounting of its effect on the radiation dose received by tumors and organs at risk during high-precision radiotherapy. Most models are parameterized by phase of the respiratory cycle and assume repeatability, although patient breathing is known to be variable. Furthermore, prior models have not considered interfraction variations which commonly occur in the thorax and the abdomen, even when images are acquired at the same respiratory phase using an external monitor [40-42]. In this article we have described a model to estimate 3-dimensional motion in patient CT images. The proposed model does not assume repeatable breath cycles, but is parameterized by the temporal variation in diaphragm position, which in general may exhibit cycle-to-cycle as well as interfraction variations.

Based on the application of this model to lung cancer patients, we make several observations. First, two principal components appear to be sufficient to accurately model 3D respiration-induced motion in the thorax. The model requires at least one surrogate signal for each principal component; therefore, respiratory motion of any voxel in the thorax is parameterized by the voxel's location and two surrogate signals. The preliminary results suggest that the choice of current diaphragm position and its prior position approximately one-third of a respiratory period as surrogate signals are appropriate to accurately characterize motion states throughout the respiratory cycle. These findings are consistent with those of Low et al [28], whose breathing motion model of anatomical features in lung is parameterized by feature location, tidal volume and airflow as measured with spirometry. Second, comparison of RCCT images on different days suggests that the motion model is applicable to modeling interfractional organ variations. In this study, the second RCCT was limited to a 9 cm region including the diaphragm; thus, we examined only tumors in proximity to the diaphragm, where correlation is more likely to be valid.

Further studies are needed to examine the model's validity for more superiorly situated tumors. Koch *et al* have observed stronger correlation of the AP motion of superior intrapulmonary vessels with chest skin movement than with abdominal movement [43], suggesting that the chest wall may be a more suitable surrogate signal in the upper thorax. We point out that the time interval between first and second RCCT was within one week and prior to the start of radiation treatments. It is likely that, although the model may possibly be accurate over several days, recalibration of the model at approximately weekly intervals will be necessary, for example, changes in tumor size or other tissue responses during treatment may affect tumor position with respect to the surrogate. Model recalibration could be accomplished by means of respiration-correlated cone-beam CT on the treatment unit [44]. Another possible limitation of the model is its validity under deeper breathing conditions that lie outside the motion range of the RCCT used to calibrate the model. Further accrual of patient data at multiple sessions will enable investigation of this question.

The accuracy of the model depends, in turn, on the accuracy of the deformable registration algorithm. Although the purpose of the principal component analysis is to remove noise in the time-dependent displacement fields generated by deformable registration, it will not remove large inaccuracies in the deformation. For example, deformable image registration may not accurately represent the sharp changes in deformation at the boundary between moving lung and stationary chest wall [45]. Such an inaccuracy will be present in all deformation fields and thus PCA will not eliminate it. Validation of the accuracy of the deformable registration algorithm is an important prerequisite to application of the model [39].

In this study, the time interval between the two surrogate signals, (approximately one-third of a respiratory period, or $\Delta t=3$ in Eq. 9) is chosen large enough so that some difference in diaphragm position can be detected, for the purpose of distinguishing between the inspiration and expiration portions of the respiratory cycle. Diaphragm differences at $\Delta t=1$ or $\Delta t=2$ may be too small, making it difficult to distinguish end inspiration from end expiration in some patients. At intervals of approximately one-half breathing cycle ($\Delta t=4$ to 6), there may be ambiguity in distinguishing mid-inspiration from mid–expiration states.

Current study of the model's ability to predict interfraction motion states was limited to 4 cases, thus the results suggest but do not confirm the model's performance in this respect. Our plans to further test the model are to acquire respiration-correlated cone-beam CT images at treatment for a larger group of patients.

The results presented here suggest that the proposed method is a potentially useful tool for predicting respiration-induced variations in a patient's 3D images during simulation and treatment, including changes in breathing pattern, when diaphragm position is monitored. Such a tool is applicable to treatment planning and evaluation of treatment delivery. As we have shown, one potential application is to reduce artifacts in RCCT images by replacement with model-generated images. Another application to treatment planning is to use the model in combination with fluoroscopy at simulation, from which cycle-to-cycle variations in diaphragmatic motion are converted to variations in 3D motion state. This provides a means to test the sensitivity of the treatment plan to breathing variations prior to treatment, without the assumption of repeatable breathing cycles inherent in RCCT. A third application is to obtain a more accurate calculation of dose to moving organs during treatment. The availability of kilovoltage imaging systems in the treatment room will make it possible for fluoroscopic or cine (~1 Hz) acquisition of kV images concurrently with delivery of treatment. Monitoring of diaphragm position in such images, in combination with the proposed model, can provide a complete record of tumor and normal organ motion over the duration of treatment.

**Figure Captions:**

Fig.1:    Spectrum of eigenvalues and cumulative sum of eigenvalues (sum normalized to

1) for a patient.

Fig. 2:    Left column: Red-blue overlays of actual CT axial (upper) and coronal (lower) images at end expiration (EE, blue enhanced) and end inspiration (EI, red enhanced). Blue and red regions indicate areas of tissue density mismatch. Tumor is indicated by arrows. Second column: red-blue overlay of actual CT image at EI and model reconstructed image at EI. Third and fourth columns: red-blue overlays of actual CT and model reconstructed images at mid-inspiration, and mid-expiration, respectively.

Fig 3    Comparison of model-predicted and observed GTV centroid displacements relative to end-expiration, for the same RCCT session of 4 patients. Data are shown at end-inspiration (time point 0), mid-expiration (time point 3) and mid-

inspiration (time point 8), in the left-right (LR), anterior-posterior (AP) and superior-inferior (SI) directions.

Fig 4    2D polar map of average model-observer (a,c) and inter-observer (b,d) GTV surface differences, for two patients.

Fig 5    Surface display of the displacement (3D vector length) of lung voxels from end expiration to end inspiration. Left: predicted displacement from deformable image registration. Middle: model prediction using two principal components. Right: difference between the displacement of deformable registration and the model prediction. Numbers are in mm.

Fig.6:   Left: 90% confidence limit (CL) differences (3D vector length) between model-predicted and actual delineated lung surfaces at end inspiration (EI), mid-inspiration (MI) and mid-expiration (ME). Right: 2D polar map of surface differences between model prediction and actual lung of patient 2, at EI motion state.

Fig. 7:  Comparison between original respiration-correlated CT images of 3 patients at mid-inspiration (left column) and model-generated images at the same respiration state (right column) using two principal components. Arrows in the original images indicate artifacts caused by respiration-correlated sorting of CT slices, which are reduced in the model-generated images.

Fig. 8:  Red-blue overlay of the model-predicted image at end inspiration (red) and the actual image from a second RCCT (blue), for two patients. The second RCCT is acquired on a different day from the first RCCT used to calibrate the model.

Fig 9:   Comparison, between model prediction and observer, of GTV centroid displacement in the second (different day) RCCT relative to the end expiration position in the first RCCT, for different patients. Upper and lower rows correspond to end expiration and end inspiration, respectively. Left, middle and right columns show displacement in the left-right, anterior-posterior, and superior-inferior directions; positive values are in the right, anterior and superior directions.

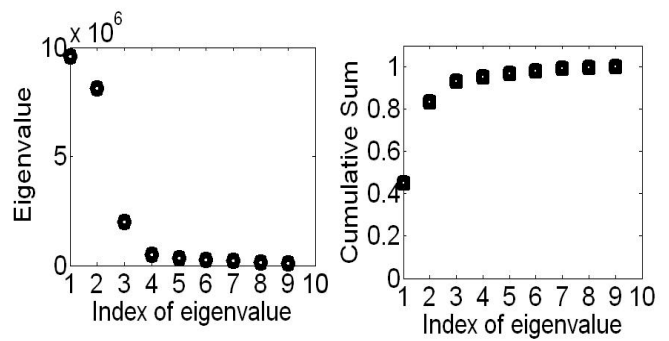

Fig.1

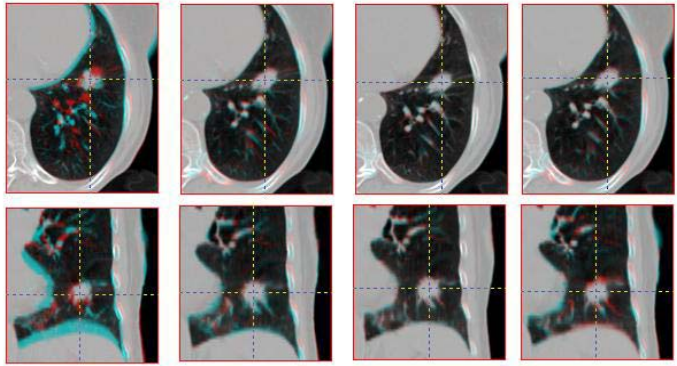

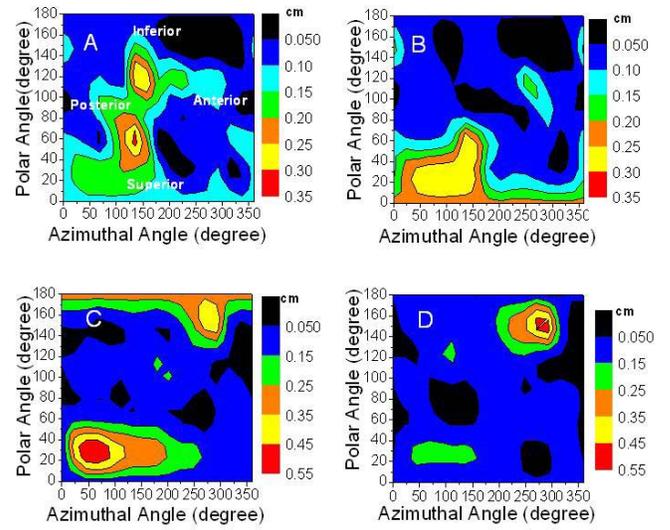

Fig. 2

Fig. 4

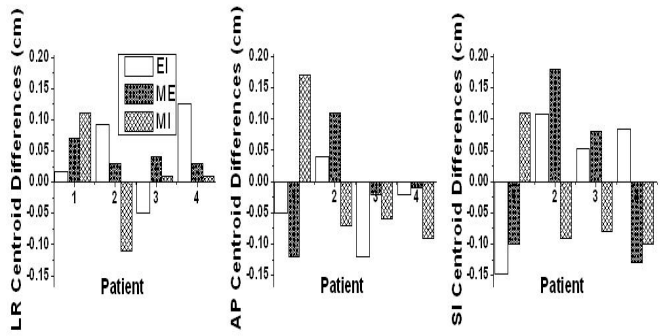

*Fig 3*

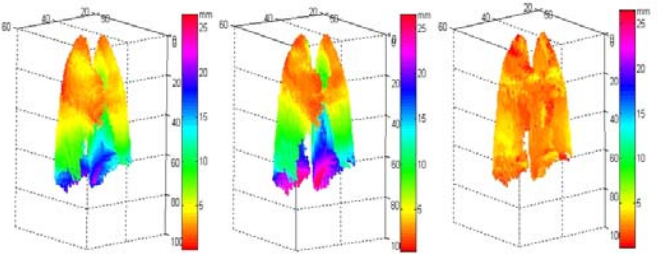

Fig. 5

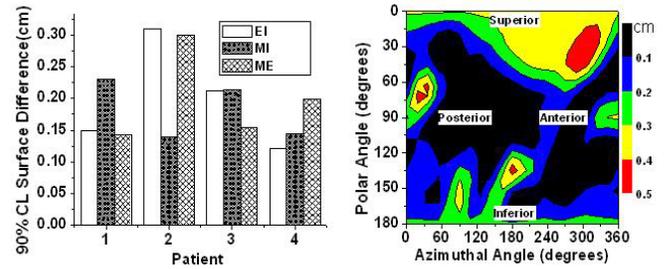

Fig.6

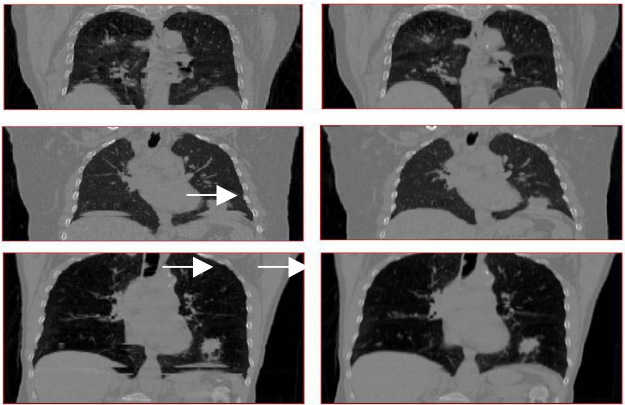

Fig. 7

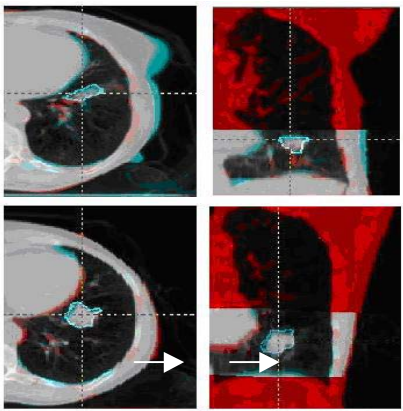

Fig.8

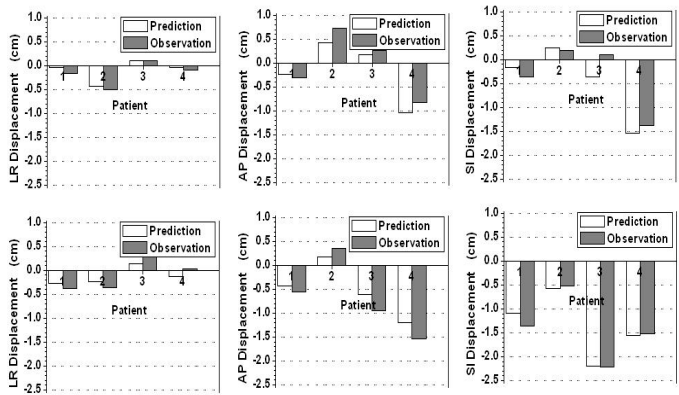

Fig. 9